\documentclass{article}

\usepackage{arxiv}

\usepackage[utf8]{inputenc} 
\usepackage[T1]{fontenc}    
\usepackage{hyperref}       
\usepackage{url}            
\usepackage{booktabs}       
\usepackage{amsfonts}       
\usepackage{amsmath}       
\usepackage{nicefrac}       
\usepackage{float}
\usepackage{microtype}      
\usepackage{cleveref}       
\usepackage{lipsum}         
\usepackage{graphicx}
\usepackage{siunitx}
\usepackage[natbibapa]{apacite}
\usepackage{doi}

\title{Handling nonlinearities and uncertainties of fed-batch cultivations with difference of convex functions tube MPC}

\newif\ifuniqueAffiliation
\uniqueAffiliationtrue

\ifuniqueAffiliation 
\author{ \href{https://orcid.org/0000-0003-2325-6001}{\includegraphics[scale=0.06]{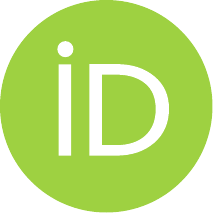}\hspace{1mm}Niels Krausch}\\
	Technische Universität Berlin\\
	Bioprocess engineering\\
	ACK24, Ackerstr. 76, 13355 Berlin, Germany\\
	\texttt{n.krausch@tu-berlin.de} \\
	\And
	\href{https://orcid.org/0000-0001-6416-3026}{\includegraphics[scale=0.06]{orcid.pdf}\hspace{1mm}Martin Doff-Sotta} \\
	University of Oxford\\
	Department of Engineering Science\\
	Parks Road, Oxford, UK\\
	\texttt{martin.doff-sotta@eng.ox.ac.uk}\\
 	\And
	\href{https://orcid.org/0000-0003-2189-7876}{\includegraphics[scale=0.06]{orcid.pdf}\hspace{1mm}Mark Cannon}\thanks{Corresponding author} \\
	University of Oxford\\
	Department of Engineering Science\\
	Parks Road, Oxford, UK\\
	\texttt{mark.cannon@eng.ox.ac.uk}\\
 	\And
	\href{https://orcid.org/0000-0002-1214-9713}{\includegraphics[scale=0.06]{orcid.pdf}\hspace{1mm}Peter Neubauer} \\
	Technische Universität Berlin\\
	Bioprocess engineering\\
	ACK24, Ackerstr. 76, 13355 Berlin, Germany\\
	\texttt{peter.neubauer@tu-berlin.de} \\
 	\And
	\href{https://orcid.org/0000-0001-9461-4414}{\includegraphics[scale=0.06]{orcid.pdf}\hspace{1mm}Mariano Nicolas Cruz Bournazou} \\
	Technische Universität Berlin\\
	Bioprocess engineering\\
	ACK24, Ackerstr. 76, 13355 Berlin, Germany\\
	\texttt{mariano.n.cruzbournazou@tu-berlin.de} \\
}
\else
\usepackage{authblk}

\newbox{\orcid}\sbox{\orcid}{\includegraphics[scale=0.06]{orcid.pdf}} 

\fi

\renewcommand{\headeright}{Preprint}
\renewcommand{\shorttitle}{DC-TMPC for cultivations}

\hypersetup{
pdftitle={Handling nonlinearities and uncertainties of fed-batch cultivations with difference of convex functions tube MPC},
pdfsubject={q-bio.NC, q-bio.QM},
pdfauthor={N. Krausch},
pdfkeywords={Robust tube MPC, Data-driven control, Convex optimization, Bioprocesses}}

\begin{document}
\maketitle

\begin{abstract}
	Bioprocesses are often characterized by nonlinear and uncertain dynamics. This poses particular challenges in the context of model predictive control (MPC). Several approaches have been proposed to solve this problem, such as robust or stochastic MPC, but they can be computationally expensive when the system is nonlinear. Recent advances in optimal control theory have shown that concepts from convex optimization, tube-based MPC, and difference of convex functions (DC) enable stable and robust online process control. The approach is based on systematic DC decompositions of the dynamics and successive linearizations around feasible trajectories. By convexity, the linearization errors can be bounded tightly and treated as bounded disturbances in a robust tube-based MPC framework. However, finding the DC composition can be a difficult task. To overcome this problem, we used a neural network with special convex structure to learn the dynamics in DC form and express the uncertainty sets using simplices to maximize the product formation rate of a cultivation with uncertain substrate concentration in the feed. The results show that this is a promising approach for computationally tractable data-driven robust MPC of bioprocesses.
\end{abstract}


\section{Introduction}
\subsection{Rapid bioprocess development}
The accelerating demand for cost-effective production of biologic drugs and sustainable biomaterials intensifies the need for rapid bioprocess development. This is particularly true in the early project stages, characterized by limited process information and a broad spectrum of potential optimal conditions. Advanced control approaches like MPC coupled with online parameter estimation of the model have proven to be successful even when incomplete process information is available \citep{Krausch.2022}, but have been restricted to relatively stable process conditions. For example, \citet{Kager.2020} were able to increase total product formation in a fungal process, but their approach is limited to the nominal case. \citet{Mowbray.2022} used Neural Networks (NN) to deal with uncertainties but required heavy offline training. 

\subsection{Tube-based MPC with difference of convex functions}
A popular approach in advanced control to deal with uncertain dynamic systems is tube-based MPC (TMPC). TMPC has been mainly applied to linear systems because nonlinear robust MPC requires online solution of nonconvex optimization problems, which can be computationally expensive. A common strategy for applying TMPC to nonlinear systems is to treat the nonlinearity as bounded disturbances of the system and perform successive linear approximations around predicted trajectories. These approaches, nevertheless, rely on conservative estimates of the linearization error and can lead to poor performance \citep{Yu.2013}. 
Recent studies have shown that tighter bounds on the linearization error can be achieved if the problem can be expressed as a difference of convex functions \citep{DoffSotta.2022b}. This is based on the observation that the necessarily convex linearization error is maximum at the boundary of the set on which it is evaluated. Tight bounds can thus be derived and treated as disturbances in a robust TMPC framework. Moreover, the DC structure of the dynamics is attractive as it results in a sequence of convex programs that can be solved with predictable computational effort. 
Even though any twice continuously differentiable function can be expressed in DC form, finding such functions can be a difficult task. To solve this problem, we have harnessed an NN by restricting the kernel weights to non-negative values and used a convex activation function (ReLU) leading to a so-called input-convex NN (ICNN) \citep{Amos.2017}. Two ICNNs can thus be stacked and their output subtracted to learn the dynamics of the function in DC form \citep{Sankaranarayanan.2022}. Moreover, in the context of TMPC, the parameterization of the tube plays an important role in the computational complexity of the optimization problem. \citet{DoffSotta.2022b} propose state tube cross sections parameterized by elementwise bounds, yielding $2^{n_x+1}$ (with $n_x$ the number of states) inequality constraints and causing a significant computational burden for large number of states. In this regard, using simplex tubes is a computationally efficient alternative with only $n_x+1$ inequality constraints. 
Hence, this contribution describes a TMPC algorithm leveraging a NN for learning the dynamics in DC form, implementing a simplex tube and optimizing product formation in a case study of a fed-batch bioreactor for the production of penicillin.

\section{Modelling and DC approximation with neural networks}
\label{sec:headings}

Let us consider a perfectly mixed isothermal fed-batch bioreactor, a popular case study example from \citet{Srinivasan.2003}. The model states are the cell concentration X [\unit{\gram\per\liter}], product concentration P [\unit{\gram\per\liter}], substrate concentration S [\unit{\gram\per\liter}] and volume V [\unit{\liter}]. The input is the feed flow rate F [\unit{\liter\per\hour}] of S. The inlet substrate concentration $S_i \in [180,220]$ \unit{\gram\per\liter} is an uncertain parameter. The dynamics of the system are given by

\begin{equation}
\begin{gathered}
\dot{X}=\mu(S)X-\frac{F}{V}X\\
\dot{S}=-\mu(S)\frac{X}{Y_{X/S}}-\frac{vX}{Y_{P/S}}+\frac{F}{V}(S_i-S)\\
\dot{P}=vX-\frac{F}{V}P\\
\dot{V}=F
\end{gathered}
\end{equation}

where $\mu(S)=\mu_{max} \frac{S}{S+K_S+S^2/K_i}$ and $\mu_{max}$ denotes the maximal growth rate (0.02 \unit{\per\hour}), $K_S$ the affinity constant of the cells towards the substrate (0.05 \unit{\gram\per\liter}), $K_i$ an inhibition constant which inhibits growth at high substrate concentrations (5 \unit{\gram\per\liter}), v the production rate (0.004 \unit{\liter\per\hour}), $Y_{X/S}$ the yield coefficient of biomass per substrate (0.5 \unit{\gram_X\per\gram_S}) and $Y_{P/S}$ the yield coefficient of product per substrate (1.2 \unit{\gram_P\per\gram_S}). The initial conditions are $X(0)=1$ \unit{\gram\per\liter}, $S(0)=0.5$ \unit{\gram\per\liter}, $P(0)=0$ \unit{\gram\per\liter} and $V(0)=120$ \unit{\liter}.\\
An NN framework was used to approximate the nonconvex dynamics as a difference of convex functions by subtracting the outputs of two ICNN subnetworks. An ICNN with $L$ layers is characterized by a parameter set $\theta={\Theta_{1:L-1},\Phi_{0:L-1},b_{0:L-1}}$ and input-output map given by $z_L=f(y;\theta)$, defined $\forall l \in\{0,\ldots,L-1\}$ by

\begin{equation}
z_{l+1}=\sigma(\Theta_l z_l + \Phi_l x + b_l)
\end{equation}

where $y$ is the input, $z_l$ is the layer activation, $\Theta_l$ are positively constrained kernel weights $(\{\Theta_l\}_{ij})\geq 0 \forall l \in \{1,...,L-1\}$, $\Phi_l$ are input passthrough weights, $b_l$ are bias and $\sigma(\cdot)$ is a convex activation function (ReLU). Each layer of an ICNN thus consists in the  composition of a convex function with a nondecreasing convex function, which implies that $z_{l+1} = f(y;\theta)$ is convex with respect to $y$. Choosing $z_{l+1}=\dot{x}$ and $y=(x,u)$, where $x=(X,S,P,V)$ and $u=F$ are the state and input of (1), two ICNN whose outputs are subtracted can be trained simultaneously to learn  the nonconvex dynamics in (1) as a difference of (elementwise) convex functions  $f_1$, $f_2$:
\begin{equation}
\dot{x}=f_1(x,u)-f_2(x,u)
\end{equation}

\begin{figure}[H]
	\centering
	\includegraphics[width=0.6\textwidth]{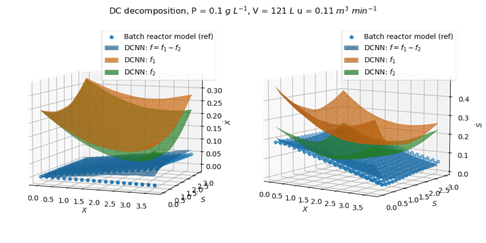}
	\caption{DC decomposition. Depicted are the results from the actual model (blue dots), the results from the DC decomposition $f=f_1-f_2$ (blue plane) and the respective DC part convex functions $f_1$ (orange) and $f_2$ (green) at a given product concentration, volume and feed rate for two states. Left: Biomass $\dot{X}$ Right: Substrate $\dot{S}$. Each in dependence of different concentrations of $X$ and $S$.}
	\label{fig:fig1}
\end{figure}

The two ICNNs each consist of a single input layer, two hidden layers with 64 nodes each and an output layer. The network was implemented in Keras and trained over 10 epochs with the RMSProp optimizer on 100,000 random samples of (1), which were divided into 80\% training and 20\% validation sets. Convexity of the models was evaluated by checking the numerical Hessian matrix of the functions for positive semidefiniteness, i.e. $\nabla^2 f_i(x,u; \theta ) \succeq 0, \forall x \in \mathbb{R}^{n_x}, \forall i = \{1,2\} $. Figure \ref{fig:fig1} depicts a 3D projection of the DC decomposition for fixed values of the states and input. As illustrated, the NN was able to obtain a good fit (MAE: 0.016) for the predictions of the ODEs (blue dots and blue surface), and the DC form of the decomposition is apparent (orange and green surfaces). 

\section{DC-TMPC framework with simplices}
\cite{DoffSotta.2022b} proposed a robust TMPC algorithm based on successive linearisation for DC systems. The so-called DC-TMPC algorithm capitalises on the idea that the successive linearisation steps yield necessarily convex linearisation error functions that can be bounded tightly and treated as disturbances by a robust MPC scheme. We extend that approach to nonconvex systems learned in DC form and consider a state tube parameterized by simplices to reduce computational burden.   
The system in DC form in (3) is discretized and successively linearized around previously computed predicted trajectories $x_k^\circ$, $u_k^\circ$ with state and input perturbations $s_k=x_k-x_k^\circ$ and $v_k=u_k-u_k^\circ$. As per the TMPC paradigm, $v_k$ is parameterized by a two degree of freedom control law $v_k=K_k$ $s_k+c_k$ where $K_k$ is a feedback gain and $c_k$ is a feedforward control sequence computed at every time step. The sequence of sets $S_k \forall s_k, \forall k$, defines the cross sections of an uncertainty tube in which the system trajectories lie under all realisations of the uncertainty and whose dynamics are given by

\begin{equation}
s_{k+1} = (\Phi_{1,k} - \Phi_{2,k})s_k + (B_{1,k} - B_{2,k})c_k + g_1(s_k,c_k x_k^\circ,u_k^\circ) - g_2(s_k,c_k x_k^\circ,u_k^\circ )
\end{equation}

where for $i=1,2$, $g_i=f_i (x_k^\circ + s_k, u_k^\circ + K_k s_k + c_k) - f_i (x_k^\circ,u_k^\circ)-\Phi_{i,k} s_k - B_{i,k} c_k$ are the (necessarily convex) linearization errors of $f_i, A_(i,k)= \frac{\partial f_i}{\partial x}  (x_k^\circ,u_k^\circ )$, $B_{i,k} = \frac{\partial f_i}{\partial u} (x_k^\circ,u_k^\circ)$ and $\Phi_{i,k} = A_{i,k} + B_{i,k} K_k$. While the approach in \cite{DoffSotta.2022b} was to parameterize the tube with elementwise bounds, resulting in an exponential increase of the inequality constraints, we consider here parameterizations of $S_k$ in terms of simplices

\begin{equation}
Q(s_k) \le \alpha_k, Q=\begin{bmatrix}
-I\\
1^T
\end{bmatrix}
\end{equation}
where $I \in \mathbb{R}^{n_x \times n_x}$ is the identity matrix, $1 \in \mathbb{R}^{n_x \times 1}$ is a vector of ones. The vector $\alpha_k \in \mathbb{R}^{(n_x+1)\times 1}$ is an optimization variable. Consequently, the state perturbation dynamics can now be expressed as $n_x+1$ inequalities as follows, combining (4) and (5)

\begin{equation}
\begin{gathered}
\max_{s \in \mathcal{V}(\mathcal{S}_k)}(-\Phi_{1,k}s-B_{1,k}c_k + f_2(x_k^\circ + s, u_k^\circ + K_ks + c_k) - f_2(x_k^\circ, u_k^\circ)) \leq [\alpha_{k+1}]_{1:n_x}\\
\max_{s \in \mathcal{V}(\mathcal{S}_k)} 1^T(f_1(x_k^\circ + s, u_k^\circ+K_ks + c_k) - f_1(x_k^\circ,u_k^\circ) - \Phi_{2,k}s-B_{2,k}c_k) \leq [\alpha_{k+1}]_{n_x+1}
\end{gathered}
\end{equation}

where the simplex vertices are $\mathcal{V}(\mathcal{S}_k) = \{-[\alpha_k ]_{1:n_x},-[\alpha_k ]_{1:n_x} + e_1 \sigma_k,\ldots,-[\alpha_k ]_{1:n_x} + e_n \sigma_k \}$, $\sigma_k=[\alpha_k ]_{n_x+1} + 1^T [\alpha_k ]_{1:n_x}$, and $e_1,\ldots,e_n$ are the standard basis vectors of $\mathbb{R}^(n_x )$. To obtain (6), we exploited the convexity of $g_i$ to obtain a tight lower bound on $\alpha_k$. Moreover, we note that (6) are convex inequalities by convexity of $f_i$ and that each maximum operation can be reduced to a discrete search over the vertices $\mathcal{V}(S_k )$ since the maximum of a convex function on a polytope occurs at one of the vertices. 

We design a TMPC controller to optimize the feedforward sequence $c_k$ and tube sets $S_k$ subject to (6) and $x_k \in \mathcal{X} \subset \mathbb{R}^{n_x}$, $u_k \in \mathcal{U} \subset \mathbb{R}^{n_u}, \forall k$. At each iteration we solve

\begin{equation}
\begin{gathered}
\min_{c,\alpha} \sum_{k=0}^{N-1} \max_{s \in \mathcal{V}(\mathcal{S}_k)} ||x_k^\circ+s-x_r||_Q^2 + \max_{s \in \mathcal{V}(\mathcal{S}_k)} ||u_k^\circ + K_ks - u_r||_R^2\\
s.t.\\
\forall k \in \{0,\ldots,N-1\}, \forall s \in \mathcal{V}(\mathcal{S}_k):
\max_{s \in \mathcal{V}(\mathcal{S}_k)}(-\Phi_{1,k}s-B_{1,k}c_k + f_2(x_k^\circ + s, u_k^\circ + K_ks + c_k) - f_2(x_k^\circ, u_k^\circ)) \leq [\alpha_{k+1}]_{1:n_x}\\
\max_{s \in \mathcal{V}(\mathcal{S}_k)} 1^T(f_1(x_k^\circ + s, u_k^\circ+K_ks + c_k) - f_1(x_k^\circ,u_k^\circ) - \Phi_{2,k}s-B_{2,k}c_k) \leq [\alpha_{k+1}]_{n_x+1}\\
x_k^\circ + s \in \mathcal{X}, u_k^0 + K_k s + c_k \in \mathcal{U} ,\alpha_0=0 
\end{gathered}
\end{equation}

with a shrinking horizon $N$. The solution from (7) is used to update the state and input guess trajectories $(x^\circ, u^\circ)$ at next iteration with 

\begin{equation}
\begin{gathered}
s_0 \leftarrow 0\\
u_k^\circ \leftarrow u_k^\circ+c_k+K_ks_k\\
s_{k+1} \leftarrow f(x_k^\circ, u_k^\circ) - x_{k+1}^\circ\\
x_{k+1} \leftarrow f(x_k^\circ, u_k^\circ)
\end{gathered}
\end{equation}

We run (7) and (8) repeatedly until $\sum_{k=0}^{N-1} ||c_k|| ^2 < \epsilon _{tol}$ or a maximum number of iterations is reached. The control input is then implemented at time $n$ by $u[n] =u_0^\circ$. At time $n+1$, we set $x_0^\circ=x[n+1]$ and the guess trajectory is updated by

\begin{equation}
\begin{gathered}
    u_k^\circ \leftarrow u_{k+1}^\circ\\
    x_{k+1}^\circ \leftarrow f(x_k^\circ, u_k^\circ)\\
    u_{N-1} \leftarrow K(x_{N-1}^\circ-x^r) + u_r\\
    x_{N}^\circ \leftarrow f(x_{N-1}^\circ, u_{N-1}^\circ)
\end{gathered}
\end{equation}

\section{Results and discussion}
\begin{figure}[H]
	\centering
    \includegraphics[width=0.8\textwidth]{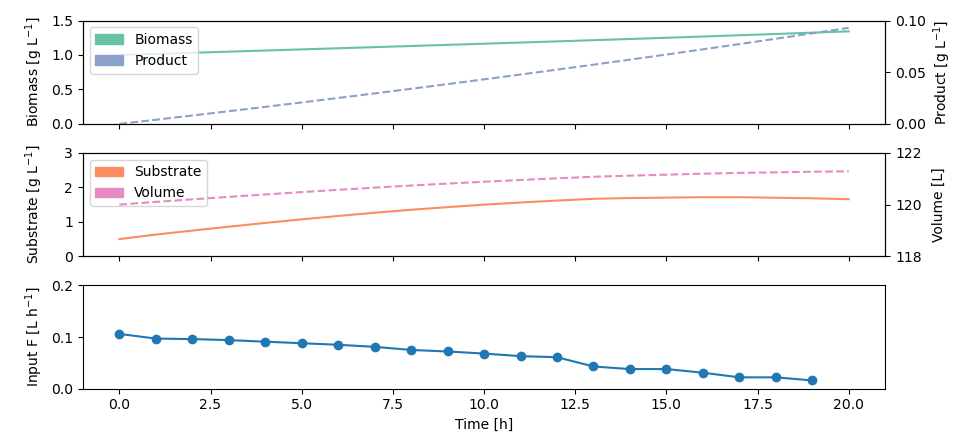}
	\caption{Results from the DC-TMPC optimization.}
	\label{fig:fig2}
\end{figure}

The proposed control algorithm was simulated on the batch reactor problem over a shrinking horizon of 20 h with a step size of 1 h using CVXPY and solver MOSEK. As shown in Figure \ref{fig:fig2}, the controller was able to maximize the product concentration with parametric uncertainty of the substrate concentration in the feed, demonstrating the applicability of this approach to complex nonlinear systems with monod-type nonlinear substrate affinity and substrate inhibition, making the search for an optimal feed rate non-trivial. The presented DC-TMPC algorithm outperforms nominal MPC approaches for this case study \citep{Lucia.2013}, by considering the uneven substrate concentration in the feed by augmenting the NN with the uncertain parameter, considering that the worst case scenario occurs at the vertices of the parameter set. Further tuning is however necessary, to find an optimal trade-off between substrate concentration in the reactor to avoid overfeeding \citep{Pimentel.2015}.

\section{Conclusion}
In this study, we show that successive linearization robust tube MPC can be an adequate tool to optimize a bioprocess under parametric uncertainty. Our approach was to decompose the nonconvex dynamics as a difference of convex functions (DC) using a neural network with convex structure and treat the necessarily convex linearization errors as bounded disturbances. Crucially, by convexity, these bounds are tight and the resulting controller is less conservative than classical TMPC based on successive linearization. This approach is relatively new and has so far only been applied to problems that already exist in DC form. Moreover, by using tubes parameterized with simplex sets, the computational effort could be significantly reduced, making it attractive for real-time optimization. Future work will incorporate more complex models and test it in a real-world bioprocess with online optimization of production in a fast-growing \textit{E. coli} strain.

\textbf{Acknowledgments}
We gratefully acknowledge the financial support of the German Federal Ministry of Education and Research (BMBF) (project no. 01DD20002A – KIWI Biolab) and the EPSRC (UKRI) Doctoral Prize scheme (grant reference number EP/W524311/1).

\bibliographystyle{apacite}
\bibliography{references}  

\begin{thebibliography}{}

\bibitem [\protect \citeauthoryear {%
Amos%
, Xu%
\BCBL {}\ \BBA {} Kolter%
}{%
Amos%
\ \protect \BOthers {.}}{%
{\protect \APACyear {2017}}%
}]{%
Amos.2017}
\APACinsertmetastar {%
Amos.2017}%
\begin{APACrefauthors}%
Amos, B.%
, Xu, L.%
\BCBL {}\ \BBA {} Kolter, J\BPBI Z.%
\end{APACrefauthors}%
\unskip\
\newblock
\APACrefYearMonthDay{2017}{}{}.
\newblock
{\BBOQ}\APACrefatitle {Input Convex Neural Networks} {Input convex neural
  networks}.{\BBCQ}
\newblock
\BIn{} D.~Precup\ \BBA {} Y\BPBI W.~Teh\ (\BEDS), \APACrefbtitle {Proceedings
  of the 34th International Conference on Machine Learning} {Proceedings of the
  34th international conference on machine learning}\ (\BVOL~70, \BPGS\
  146--155).
\newblock
\APACaddressPublisher{}{PMLR}.
\newblock
\begin{APACrefURL} \url{https://proceedings.mlr.press/v70/amos17b.html}
  \end{APACrefURL}
\PrintBackRefs{\CurrentBib}

\bibitem [\protect \citeauthoryear {%
Doff-Sotta%
\ \BBA {} Cannon%
}{%
Doff-Sotta%
\ \BBA {} Cannon%
}{%
{\protect \APACyear {2022}}%
}]{%
DoffSotta.2022b}
\APACinsertmetastar {%
DoffSotta.2022b}%
\begin{APACrefauthors}%
Doff-Sotta, M.%
\BCBT {}\ \BBA {} Cannon, M.%
\end{APACrefauthors}%
\unskip\
\newblock
\APACrefYearMonthDay{2022}{}{}.
\newblock
{\BBOQ}\APACrefatitle {Difference of convex functions in robust tube nonlinear
  MPC} {Difference of convex functions in robust tube nonlinear mpc}.{\BBCQ}
\newblock
\BIn{} \APACrefbtitle {2022 IEEE 61st Conference on Decision and Control (CDC)}
  {2022 ieee 61st conference on decision and control (cdc)}\ (\BPGS\
  3044--3050).
\newblock
\APACaddressPublisher{}{IEEE}.
\newblock
\begin{APACrefDOI} \doi{10.1109/CDC51059.2022.9993390} \end{APACrefDOI}
\PrintBackRefs{\CurrentBib}

\bibitem [\protect \citeauthoryear {%
Kager%
, Tuveri%
, Ulonska%
, Kroll%
\BCBL {}\ \BBA {} Herwig%
}{%
Kager%
\ \protect \BOthers {.}}{%
{\protect \APACyear {2020}}%
}]{%
Kager.2020}
\APACinsertmetastar {%
Kager.2020}%
\begin{APACrefauthors}%
Kager, J.%
, Tuveri, A.%
, Ulonska, S.%
, Kroll, P.%
\BCBL {}\ \BBA {} Herwig, C.%
\end{APACrefauthors}%
\unskip\
\newblock
\APACrefYearMonthDay{2020}{}{}.
\newblock
{\BBOQ}\APACrefatitle {Experimental verification and comparison of model
  predictive, PID and model inversion control in a Penicillium chrysogenum
  fed-batch process} {Experimental verification and comparison of model
  predictive, pid and model inversion control in a penicillium chrysogenum
  fed-batch process}.{\BBCQ}
\newblock
\APACjournalVolNumPages{Process Biochemistry}{90}{}{1--11}.
\newblock
\begin{APACrefDOI} \doi{10.1016/j.procbio.2019.11.023} \end{APACrefDOI}
\PrintBackRefs{\CurrentBib}

\bibitem [\protect \citeauthoryear {%
Krausch%
\ \protect \BOthers {.}}{%
Krausch%
\ \protect \BOthers {.}}{%
{\protect \APACyear {2022}}%
}]{%
Krausch.2022}
\APACinsertmetastar {%
Krausch.2022}%
\begin{APACrefauthors}%
Krausch, N.%
, Kim, J\BPBI W.%
, Barz, T.%
, Lucia, S.%
, Gro{\ss}, S.%
, Huber, M\BPBI C.%
\BDBL {}{Cruz Bournazou}, M\BPBI N.%
\end{APACrefauthors}%
\unskip\
\newblock
\APACrefYearMonthDay{2022}{}{}.
\newblock
{\BBOQ}\APACrefatitle {High-throughput screening of optimal process conditions
  using model predictive control} {High-throughput screening of optimal process
  conditions using model predictive control}.{\BBCQ}
\newblock
\APACjournalVolNumPages{Biotechnology and bioengineering}{119}{12}{3584--3595}.
\newblock
\begin{APACrefDOI} \doi{10.1002/bit.28236} \end{APACrefDOI}
\PrintBackRefs{\CurrentBib}

\bibitem [\protect \citeauthoryear {%
Lucia%
\ \BBA {} Engell%
}{%
Lucia%
\ \BBA {} Engell%
}{%
{\protect \APACyear {2013}}%
}]{%
Lucia.2013}
\APACinsertmetastar {%
Lucia.2013}%
\begin{APACrefauthors}%
Lucia, S.%
\BCBT {}\ \BBA {} Engell, S.%
\end{APACrefauthors}%
\unskip\
\newblock
\APACrefYearMonthDay{2013}{}{}.
\newblock
{\BBOQ}\APACrefatitle {Robust nonlinear model predictive control of a batch
  bioreactor using multi-stage stochastic programming} {Robust nonlinear model
  predictive control of a batch bioreactor using multi-stage stochastic
  programming}.{\BBCQ}
\newblock
\BIn{} \APACrefbtitle {2013 European Control Conference (ECC)} {2013 european
  control conference (ecc)}\ (\BPGS\ 4124--4129).
\newblock
\APACaddressPublisher{}{IEEE}.
\newblock
\begin{APACrefDOI} \doi{10.23919/ECC.2013.6669521} \end{APACrefDOI}
\PrintBackRefs{\CurrentBib}

\bibitem [\protect \citeauthoryear {%
Mowbray%
, Petsagkourakis%
, {Del Rio Chanona}%
\BCBL {}\ \BBA {} Zhang%
}{%
Mowbray%
\ \protect \BOthers {.}}{%
{\protect \APACyear {2022}}%
}]{%
Mowbray.2022}
\APACinsertmetastar {%
Mowbray.2022}%
\begin{APACrefauthors}%
Mowbray, M\BPBI R.%
, Petsagkourakis, P.%
, {Del Rio Chanona}, E\BPBI A.%
\BCBL {}\ \BBA {} Zhang, D.%
\end{APACrefauthors}%
\unskip\
\newblock
\APACrefYearMonthDay{2022}{}{}.
\newblock
{\BBOQ}\APACrefatitle {Safe chance constrained reinforcement learning for batch
  process control} {Safe chance constrained reinforcement learning for batch
  process control}.{\BBCQ}
\newblock
\APACjournalVolNumPages{Computers {\&} Chemical Engineering}{157}{}{107630}.
\newblock
\begin{APACrefDOI} \doi{10.1016/j.compchemeng.2021.107630} \end{APACrefDOI}
\PrintBackRefs{\CurrentBib}

\bibitem [\protect \citeauthoryear {%
Pimentel%
, Benavides%
, Dewasme%
, Coutinho%
\BCBL {}\ \BBA {} Wouwer%
}{%
Pimentel%
\ \protect \BOthers {.}}{%
{\protect \APACyear {2015}}%
}]{%
Pimentel.2015}
\APACinsertmetastar {%
Pimentel.2015}%
\begin{APACrefauthors}%
Pimentel, G\BPBI A.%
, Benavides, M.%
, Dewasme, L.%
, Coutinho, D.%
\BCBL {}\ \BBA {} Wouwer, A\BPBI V.%
\end{APACrefauthors}%
\unskip\
\newblock
\APACrefYearMonthDay{2015}{}{}.
\newblock
{\BBOQ}\APACrefatitle {An Observer-based Robust Control Strategy for Overflow
  Metabolism Cultures in Fed-Batch Bioreactors} {An observer-based robust
  control strategy for overflow metabolism cultures in fed-batch
  bioreactors}.{\BBCQ}
\newblock
\APACjournalVolNumPages{IFAC-PapersOnLine}{48}{8}{1081--1086}.
\newblock
\begin{APACrefDOI} \doi{10.1016/j.ifacol.2015.09.112} \end{APACrefDOI}
\PrintBackRefs{\CurrentBib}

\bibitem [\protect \citeauthoryear {%
Sankaranarayanan%
\ \BBA {} Rengaswamy%
}{%
Sankaranarayanan%
\ \BBA {} Rengaswamy%
}{%
{\protect \APACyear {2022}}%
}]{%
Sankaranarayanan.2022}
\APACinsertmetastar {%
Sankaranarayanan.2022}%
\begin{APACrefauthors}%
Sankaranarayanan, P.%
\BCBT {}\ \BBA {} Rengaswamy, R.%
\end{APACrefauthors}%
\unskip\
\newblock
\APACrefYearMonthDay{2022}{}{}.
\newblock
{\BBOQ}\APACrefatitle {CDiNN -- Convex difference neural networks} {Cdinn --
  convex difference neural networks}.{\BBCQ}
\newblock
\APACjournalVolNumPages{Neurocomputing}{495}{}{153--168}.
\newblock
\begin{APACrefDOI} \doi{10.1016/j.neucom.2022.01.024} \end{APACrefDOI}
\PrintBackRefs{\CurrentBib}

\bibitem [\protect \citeauthoryear {%
Srinivasan%
, Bonvin%
, Visser%
\BCBL {}\ \BBA {} Palanki%
}{%
Srinivasan%
\ \protect \BOthers {.}}{%
{\protect \APACyear {2003}}%
}]{%
Srinivasan.2003}
\APACinsertmetastar {%
Srinivasan.2003}%
\begin{APACrefauthors}%
Srinivasan, B.%
, Bonvin, D.%
, Visser, E.%
\BCBL {}\ \BBA {} Palanki, S.%
\end{APACrefauthors}%
\unskip\
\newblock
\APACrefYearMonthDay{2003}{}{}.
\newblock
{\BBOQ}\APACrefatitle {Dynamic optimization of batch processes} {Dynamic
  optimization of batch processes}.{\BBCQ}
\newblock
\APACjournalVolNumPages{Computers {\&} Chemical Engineering}{27}{1}{27--44}.
\newblock
\begin{APACrefDOI} \doi{10.1016/S0098-1354(02)00117-5} \end{APACrefDOI}
\PrintBackRefs{\CurrentBib}

\bibitem [\protect \citeauthoryear {%
Yu%
, Maier%
, Chen%
\BCBL {}\ \BBA {} Allg{\"o}wer%
}{%
Yu%
\ \protect \BOthers {.}}{%
{\protect \APACyear {2013}}%
}]{%
Yu.2013}
\APACinsertmetastar {%
Yu.2013}%
\begin{APACrefauthors}%
Yu, S.%
, Maier, C.%
, Chen, H.%
\BCBL {}\ \BBA {} Allg{\"o}wer, F.%
\end{APACrefauthors}%
\unskip\
\newblock
\APACrefYearMonthDay{2013}{}{}.
\newblock
{\BBOQ}\APACrefatitle {Tube MPC scheme based on robust control invariant set
  with application to Lipschitz nonlinear systems} {Tube mpc scheme based on
  robust control invariant set with application to lipschitz nonlinear
  systems}.{\BBCQ}
\newblock
\APACjournalVolNumPages{Systems {\&} Control Letters}{62}{2}{194--200}.
\newblock
\begin{APACrefDOI} \doi{10.1016/j.sysconle.2012.11.004} \end{APACrefDOI}
\PrintBackRefs{\CurrentBib}

\end{thebibliography}

\end{document}